\documentclass[pra,twocolumn,a4paper,showpacs,superscriptaddress]{revtex4}
\usepackage{amsmath}
\usepackage{amsfonts}
\usepackage{graphicx}
\begin{document}
\title{Uncertainty relations in the realm of classical dynamics}
\author{A. R. Usha Devi}
\email{arutth@rediffmail.com}
\affiliation{Department of Physics, Bangalore University, 
Bangalore-560 056, India}
\affiliation{Inspire Institute Inc., Alexandria, Virginia, 22303, USA.}
\author{H. S. Karthik}
\affiliation{Raman Research Institute, Bangalore 560 080, India}

\date{\today}

\begin{abstract}
It is generally believed that classical regime emerges as a limiting case of quantum theory. Exploring such quantum-classical correspondences in a more transparent manner is central to the deeper understanding of foundational aspects  and has attracted a great deal of attention -- starting from the early days of quantum theory. While it is often highlighted that quantum to classical transition occurs in the limit $\hbar\rightarrow 0$, several objections have been raised about its suitability in some physical contexts. Ehrenfest's theorem is another widely discussed classical limit -- however, its inadequacy has also been pointed out in specific examples. It has been proposed that since a quantum mechanical wave function inherits an intrinsic statistical behavior, its classical limit must correspond to a classical ensemble -- not an individual particle. This opens up the question ``how would uncertainty relations of canonical observables compare themselves in quantum and classical realms?" In this paper we explore parallels between uncertainty relations  in stationary states of quantum systems and that in the corresponding classical ensemble. We confine ourselves to one dimensional conservative systems and show, with the help of suitably defined dimensionless physical quantities, that first and second moments of the canonical observables match with each other in classical and quantum descriptions --  resulting in an identical structure for uncertainty relations in both the realms.       
\end{abstract}
\maketitle

\section{Introduction} 

It is imperative to retrieve classical dynamics as a limiting case -- in its domain of validity -- from quantum theory.
 The generally prevailing notion is  that classical mechanics emerges in the limit $\hbar\rightarrow 0$. Yet another classical limit discussed widely is the one following Ehrenfest's theorem. Pedagogic discussions, in several text books on quantum mechanics, are essentially confined to these two approaches towards classical regime.  However, both these routes are shown to be not universally satisfactory~\cite{Berry, Nemes, Ballentine1,Sen, Ballentine2, Angelo1, Angelo2}. It has been pointed out that  classical realm -- resulting from a quantum mechanical state -- is ought to correspond to an ensemble -- not a  single particle~\cite{Ballentine1, Huang}. The averages, variances and other higher order moments of the quantum and classical probability distributions are therefore expected to agree in the limiting case. 
 
In order to compare the statistical form of classical dynamics with the corresponding one in quantum dynamics, phase space probability distribution of the classical ensemble (a counterpart of the corresponding quantum state) needs to be identified. The classical phase space probability distribution satisfies the  Liouville equation and  the phase space  averages of the classical observables are shown to exhibit analogous dynamical behaviour as that of the corresponding quantum case -- even when Ehrenfest's theorem breaks down~\cite{Ballentine1}. 
 
Another approach, when one confines to stationary state solutions of the quantum Hamiltonian, is to graphically compare the probability density function $P^{(n)}_{\rm QM}(x)=\vert \psi_n(x)\vert^2$ with the corresponding classical probability distribution $P_{\rm CL}(x)$ of an ensemble and to recognize that the {\em envelope} of the quantum probability density approaches the classical one  in the  large $n$ limit~\cite {Rob}. 

In this paper, we show that the first and second moments of suitably defined {\em dimensionless} canonical variables evaluated in the stationary states of one dimensional conservative  quantum systems match  with those associated with the corresponding classical ensemble. 
 This, in turn leads to  {\em identical} structure for uncertainty relations of the dimensionless position and momentum variables in both classical and quantum domains -- bringing out the underlying unity of the two formalisms -- irrespective of their structually different mathematical and conceptual nature.

\section{Classical probability distributions corresponding to quantum mechanical stationary states}

We begin here by reviewing the classical probability distributions~\cite{Rob} for an ensemble of particles bound in one dimensional potentials $V(x)$. 
The probability density function for position of a {\em single} particle, whose initial position and velocities are specified, is given by
\begin{equation}
\label{1}
P^{\rm single}_{\rm CL}(x)=\delta[x-x(t)] 
\end{equation} 
where $x(t)$ denotes the deterministic trajectory of the particle at any instant of time $t$. However, the quantum mechanical  probability density $P^{(n)}_{QM}(x)=\vert\psi_n(x)\vert^2$ associated with the stationary state solution $\psi_n(x)$  of the system, is not expected to approach -- in the classical realm -- to  the single particle probability density of Eq.~(\ref{1})  --- rather,  the locally averaged quantum probability density 
{\em does} approximate to a probability distribution $P_{\rm CL}(x)$ of a classical ensemble of particles (of fixed energy $E$) in the large $n$ limit~\cite{Rob}. 

The  phase space probability distribution $P_{\rm CL}(x,p)$ of an ensemble of classical particles of fixed energy $E$ and bound in a potential $V(x)$     is  proportional to $\delta\left[\frac{p^2}{2m}+V(x)-E\right]$. The position probability function is obtained by integrating over the momentum variable $p$:
\begin{eqnarray}
\label{posprob}
P_{\rm CL}(x)&=&\, \int\, dp\, P_{\rm CL}(x,p)  \nonumber \\
&=& {\rm Constant}\cdot \, \int\, dp\, \delta\left[\frac{p^2}{2m}+V(x)-E\right].
\end{eqnarray} 
Using the properties $\delta(\alpha\, x)=\frac{1}{\alpha}\, \delta(x)$ and $\delta (x^2-a^2)=\frac{1}{2a}\left[\delta(x+a)+\delta(x-a)\right]$ of the Dirac delta function, the classical probaility distribution reduces to
\begin{widetext}
\begin{eqnarray}
\label{posprob2}
P_{\rm CL}(x)&=&{\rm Constant}\cdot \, \int dp\, 2m\, \delta\left(p^2+2m[V(x)-E]\right) \nonumber \\
&=&  {\rm Constant}\cdot \,  \sqrt{\frac{2m}{[E-V(x)]}}\,  \int dp\, \, \left[\delta\left(p+\sqrt{2m[E-V(x)]}\right)+\delta\left(p-\sqrt{2m[E-V(x)]}\right)\right]\nonumber \\ 
&=&  \frac{{\cal N}}{\sqrt{E-V(x)}},
\end{eqnarray} 
\end{widetext}
where ${\cal N}$ denotes the normalization factor, such that $\int_{-A}^{A}\, dx\, P_{\rm CL}(x)=1$ (here the integration is taken between $(-A,A)$ as the probability distribution $P_{\rm CL}(x)$ vanishes outside the classical turning points $\vert x\vert > A$).  It may be readily seen that, by substituting $E=\frac{1}{2}m\omega^2\,A^2$, and $V(x)=\frac{1}{2}m\omega^2\,x^2$, in the familiar example of  harmonic oscillator, the position probability distribution Eq.~(\ref{posprob2}) of the classical ensemble reduces to the well-known expression $P_{\rm CL}(x)=\frac{1}{\pi}\, \frac{1}{\sqrt{A^2-x^2}}$.     

The phase space averages of any arbitary function $F(x,p)$ of position and momentum variables get reduced to those evaluated with position probability distribution function $P_{\rm CL}(x)$ as follows:  
\begin{widetext}
\begin{eqnarray}
\label{fxp}
\langle F(x,p)\rangle_{\rm CL} &=& \int\, dx \int dp\, P_{\rm CL}(x,p)\, F(x,p)\nonumber \\
 &=&  {\rm Constant}\cdot \, \int\, dx  \,  \int dp\,  \delta\left(\frac{p^2}{2m}+V(x)-E\right)\, F(x,p) \nonumber \\
 &=& {\rm Constant}\cdot \, \int\, dx \, \sqrt{\frac{2m}{E-V(x)}}\,  \int dp\, \,  \left[ \delta\left(p+\sqrt{2m[E-V(x)]}\right) +\delta\left(p-\sqrt{2m[E-V(x)]}\right)\right]\, F(x,p) \nonumber \\
&=& {\rm Constant}\cdot \, \int\, dx\,  \sqrt{\frac{2m}{E-V(x)}}\, \left[F(x, -\sqrt{2m[E-V(x)]}) + F(x, \sqrt{2m[E-V(x)]})\right] \nonumber \\
&=& \frac{1}{2}\, \int \, dx\, P_{\rm CL}(x)\, \left[ F\left(x, -\sqrt{2m[E-V(x)]}\right) + F\left(x, \sqrt{2m[E-V(x)]}\right)\right] 
\end{eqnarray} 
\end{widetext}

We define dimensionless (scaled) position and momentum variables  as follows, 
\begin{equation}
\label{dl}
X=\frac{x}{A}, \ \ P=\frac{p}{\sqrt{2mE}}, 
\end{equation}
such that $\vert X\vert,\, \vert P\vert \leq  1$ in a bounded system.

In the next section, we compute the first and second  moments $\langle X\rangle_{\rm CL}, \langle X^2\rangle_{\rm CL}, \langle P \rangle_{\rm CL}, \langle P^2\rangle_{\rm CL}$ of the classical probability distribution in three specific examples of one-dimensional bound systems. We then compare these classical averages with the quantum expectation values   $\langle \hat X\rangle_{\rm QM}, \langle \hat{X}^2\rangle_{\rm QM}, \langle \hat{P} \rangle_{\rm QM}, \langle \hat{P}^2\rangle_{\rm QM}$ -- evaluated in the stationary states $\psi_n(x)$ and identify that they agree with each other in the classical limit.  

\section{Comparision of first and second moments of the classical distribution with the stationary state quantum moments}   

We focus now on three specific examples of one dimensional bound systems viz.,  harmonic oscillator, infinite well  and the bouncing ball  --  both in the classical and quantum domain. We evaluate first and second  moments of dimensionless position and momentum variables (see Eq.~(\ref{dl}) for definition) and identify that the quantum moments -- evaluated in stationary eigen states of the Hamiltonian -- match with their classical counterparts.   

\subsection{One dimensional harmonic oscillator} 

The classical probability density for finding a system of harmonic oscillators -- all having same amplitude $A$ --  between  position  $x$ and $x+dx$ is given by (see the paragraph following Eq.~(\ref{posprob2}))  
\begin{equation}
\label{prho}
P_{\rm CL}(x)=\left\{\begin{array}{ll} \frac{1}{\pi}\, \frac{1}{\sqrt{A^2-x^2}}, & {\rm for} \ \vert x\vert \leq A \\
=0,  &  {\rm for} \  \vert x\vert >A. \end{array} \right.        
\end{equation} 

We consider scaled canonical variables $X=\frac{x}{A}$ and $P=\frac{p}{\sqrt{2mE}}= \frac{p}{m\, \omega\, A}$ and evaluate 
the averages of $X,\, X^2,\, P,\, P^2$  (by making use of (\ref{fxp}) and (\ref{prho})) as given below : 
\begin{widetext}
\begin{eqnarray}
\label{xho}
\langle X\rangle_{\rm CL}&=& \frac{1}{A}\, \int\, dx\, P_{\rm CL}(x)\, x 
 =\frac{1}{A\, \pi}\, \int_{-A}^{A}\,  dx\, \frac{x}{\sqrt{A^2-x^2}} =0, \\ 
\label{xsqho}
\langle X^2\rangle_{\rm CL}&=&\frac{1}{A^2}\, \int\, dx\, P_{\rm CL}(x)\, x^2\, 
 =\frac{1}{A^2\, \pi}\, \int_{-A}^{A}\,  dx\, \frac{x^2}{\sqrt{A^2-x^2}} = \frac{1}{2},  \\ 
\label{pho} 
\langle P\rangle_{\rm CL}&=& \frac{1}{2\, m\omega\, A}\, \int_{-A}^{A}\, dx\,  P_{\rm CL}(x)\,
\left(-\sqrt{2m\left[E-\frac{1}{2}m\omega^2\, x^2\right]}+\sqrt{2m\left[E-\frac{1}{2}m\omega^2\, x^2\right]}\right)    
= 0, \\
\label{psqho} 
\langle P^2\rangle_{\rm CL}&=& \frac{1}{m^2\omega^2\, A^2}\, \int_{-A}^{A}\, dx\,  P_{\rm CL}(x)\,2m\, \left[E-\frac{1}{2}m\omega^2\, x^2\right]=  \frac{1}{A^2\, \pi}\, \int_{-A}^{A}\,  dx\, \sqrt{A^2-x^2} = \frac{1}{2}. 
\end{eqnarray} 
\end{widetext}
The variances of $X$, $P$ are given by, 
\begin{eqnarray}
\label{varxpho} 
 (\Delta X)_{\rm CL}^2&=&\langle X^2\rangle_{\rm CL}-\langle X\rangle_{\rm CL}^2=\frac{1}{2} \nonumber \\
 (\Delta P)_{\rm CL}^2&=&\langle P^2\rangle_{\rm CL}-\langle P\rangle_{\rm CL}^2=\frac{1}{2}
\end{eqnarray}
and hence the product of variances obey, 
\begin{equation}
\label{unho}
 (\Delta X)_{\rm CL}^2\, (\Delta P)_{\rm CL}^2\equiv\frac{1}{4} 
\end{equation}
in a classical ensemble (characterized by the probability distribution (\ref{prho})) of harmonic oscillators. 

The stationary state solutions of the quantum Hamiltonian
\begin{equation}
\hat{H}=\frac{\hat{p}^2}{2m}+\frac{1}{2}\, m\,\omega^2\, \hat{x}^2 
\end{equation}  
are given by, 
\begin{equation}
\label{host}
\psi_n(x)=\left(\frac{\sqrt{m\omega/\pi\hbar}}{2^n\, n!}\right)^{1/2}\, H_n(\sqrt{m\omega/\hbar}\, x)\, e^{-\frac{m\omega\, x^2}{\hbar}} 
\end{equation}  
where $H_n$ are Hermite polynomials of degree $n$; the corresponding energy eigen values are, 
\begin{equation}
E_n=\left(n+\frac{1}{2}\right)\, \hbar\omega,\ n=0,1,2,\ldots.
\end{equation}
The {\em classical turning points} associated with the energy $E_n$ are readily identified as, 
\begin{equation}
A_n=\sqrt{\frac{2\, E_n}{m\omega^2}}=\sqrt{\frac{(2n+1)\, \hbar}{m\omega}}. 
\end{equation}
We consider scaled position and momentum operators, 
\begin{eqnarray} 
\hat{X}&=&\frac{\hat{x}}{A_n}=\hat{x}\, \sqrt{\frac{m\omega}{(2n+1)\,\hbar}}  \nonumber \\ 
\hat{P}&=&\frac{\hat{p}}{\sqrt{2\,m\,E_n}}= \frac{ \hat{p}}{\sqrt{(2n+1)\, \hbar\, m\,\omega}}
\end{eqnarray} 
corresponding to the classical ones $X=\frac{x}{A},\ P=\frac{p}{\sqrt{2mE}}$. 
The expectation values of $\hat{X},\, \hat{X}^2, \hat{P}$, and $\hat{P}^2$, evaluated in the stationary states $\psi_n(x)$ (see Eq.~(\ref{host})) are given by, 
\begin{widetext}
\begin{eqnarray}
\label{qhox}
\langle \hat{X}\rangle_{QM}&=&\sqrt{\frac{m\omega}{(2n+1)\,\hbar}}\, \int_{-\infty}^{\infty} \, dx\, \vert \psi_n(x)\vert^2\, x \nonumber \\
&=&\, \frac{m\omega}{\hbar\, 2^n\, n!\, \sqrt{(2n+1)\, \pi}}\, \int_{-\infty}^{\infty} \, dx\, H_n^2(\sqrt{m\omega/\hbar}\, x)\, e^{-\frac{2\,m\omega\, x^2}{\hbar}}\, x =0, \\
\label{qhoxsq}
\langle \hat{X^2}\rangle_{QM}&=&\frac{m\omega}{(2n+1)\,\hbar}\, \int_{-\infty}^{\infty} \, dx\, \vert \psi_n(x)\vert^2\, x^2 \nonumber \\
&=&\, \left(\frac{m\omega}{\hbar}\right)^{3/2}\, \frac{1}{2^n\, n!\, (2n+1)\, \sqrt{\pi}}\, \int_{-\infty}^{\infty} \, dx\, H_n^2(\sqrt{m\omega/\hbar}\, x)\, e^{-\frac{2\,m\omega\, x^2}{\hbar}}\, x^2 =\frac{1}{2}, \\
\label{qhop}
\langle \hat{P}\rangle_{QM}&=&-i\sqrt{\frac{\hbar}{(2n+1)\, m\,\omega}}\, \int_{-\infty}^{\infty} \, dx\,  \psi^*_n(x)\frac{d\psi_n(x)}{dx} \nonumber \\
&=&\frac{-i}{2^n\, n!\, \sqrt{(2n+1)\, \pi}}\, \int_{-\infty}^{\infty} \, dx\, H_n(\sqrt{m\omega/\hbar}\, x)\, e^{-\frac{\,m\omega\, x^2}{\hbar}}
\, \frac{d}{dx}\, \left(H_n(\sqrt{m\omega/\hbar}\, x)\, e^{-\frac{\,m\omega\, x^2}{\hbar}}\right) =0, \\
\label{qhopsq}
\langle \hat{P^2}\rangle_{QM}&=&\frac{-\hbar}{(2n+1)\, m\,\omega}\, \int_{-\infty}^{\infty} \, dx\,  \psi^*_n(x)\frac{d^2\psi_n(x)}{dx^2} \nonumber \\
&=&-\sqrt{\frac{\hbar}{m\omega\, \pi}}\, \frac{1}{2^n\, n!\, (2n+1)}\, \int_{-\infty}^{\infty} \, dx\, H_n(\sqrt{m\omega/\hbar}\, x)\, e^{\frac{-m\omega\, x^2}{\hbar}}
\, \frac{d^2}{dx^2}\, \left(H_n(\sqrt{m\omega/\hbar}\, x)\, e^{-\frac{\,m\omega\, x^2}{\hbar}}\right)=\frac{1}{2}.
\end{eqnarray}
\end{widetext}  
Clearly, the quantum expectation values Eqs.(\ref{qhox})-(\ref{qhopsq}) match with the classical ones given in Eqs.(\ref{xho})-(\ref{psqho}) and  we obtain the uncertainty product -- for {\em all} stationary state solutions of the quantum oscillator, 
\begin{equation}
\label{qhoun} 
\left(\Delta\, \hat{X}\right)_{\rm QM}\,\left(\Delta\, \hat{P}\right)_{\rm QM}\equiv \frac{1}{4}.  
\end{equation} 

It is pertinent to point out here that the commutator relation, 
\begin{eqnarray}
[\hat{X},\hat{P}]&=&\left[\sqrt{\frac{m\omega}{(2n+1)\,\hbar}} \hat{x}, \frac{ \hat{p}}{\sqrt{\hbar\, m\,\omega\, (2n+1)}}\right] \nonumber \\ 
&=&\frac{1}{(2n+1)\, \hbar}\, [\hat{x},\hat{p}]=\frac{i}{2n+1},
\end{eqnarray} 
leads to  the uncertainty relation, 
\begin{eqnarray}
 \left(\Delta\, \hat{X}\right)_{QM}^2\, \left(\Delta\, \hat{P}\right)_{QM}^2\geq \frac{1}{4\, (2n+1)^2}.
\end{eqnarray}
In the large $n$ limit one obtains the result $\lim_{n\rightarrow \infty}\, \left(\Delta\, \hat{X}\right)_{QM}^2\, \left(\Delta\, \hat{P}\right)_{QM}^2\geq 0$ -- which is usually expected in the classical regime. However, the exact result (\ref{qhoun}) for uncertainty product holds for {\em all} the stationary state solutions  -- and strikingly, this result matches with that  of  a classical ensemble of oscillators with fixed energy $E$  (see (\ref{unho})). 

\subsection{One dimensional infinite potential box}

We consider a symmetric infinite potential well defined by, 
\begin{equation}
\label{infpot}
V(x)=\left\{\begin{array}{ll} 0 & {\rm for}\ -\frac{L}{2}\leq x\leq \frac{L}{2}, \\ 
\infty & {\rm for}\ \  \vert x\vert > \frac{L}{2}.\end{array}\right. 
\end{equation} 
The particles  move with a constant velocity within the box and get reflected back and forth (i.e., the momentum changes sign when the particles hit the walls). The  position probability distribution for an ensemble of classical particles confined to move within the box (such that $-\frac{L}{2}\leq x\leq \frac{L}{2}$) is a constant (as can be readily seen by substituting Eq.~(\ref{infpot}) in Eq.~(\ref{posprob2})  and is given by~\cite{Rob}, 
\begin{equation}
\label{iwprob}
P_{\rm CL}(x)=\left\{\begin{array}{ll} \frac{1}{L} & {\rm for}\  \vert x\vert \leq \frac{L}{2} \\
0 & {\rm for}\  \vert x\vert > \frac{L}{2}
\end{array}\right., 
\end{equation}
which obeys $\int_{-L/2}^{L/2}\, P_{\rm CL}(x)\, dx=1$. 

In this example, the dimensionless position and momentum variables are identified as, 
\begin{equation}
\label{iwXP}
X=\frac{x}{(L/2)},\ \ P=\frac{p}{\sqrt{2mE}}=\frac{p}{\vert p\vert}
\end{equation}
and the classical averages  $\langle X\rangle_{\rm CL}, \langle X^2\rangle_{\rm CL}, \, \langle P\rangle_{\rm CL}, \langle P^2\rangle_{\rm CL}$ are readily  evaluated using the probability distribution (\ref{iwprob}):  
\begin{eqnarray}
\label{xiw}
\langle X\rangle_{\rm CL}&=&\int\, dx\, P_{\rm CL}(x)\, \frac{x}{L/2}\nonumber \\ 
&=&\frac{2}{L^2}\, \int_{-L/2}^{L/2}\, x dx =0, \\ 
\label{xsqiw}
\langle X^2\rangle_{\rm CL}&=&\int\, dx\, P_{\rm CL}(x)\, \frac{x^2}{L^2/4}\nonumber \\
&=&\frac{4}{L^3}\, \int_{-L/2}^{L/2}\,  dx\, x^2 = \frac{1}{3},  \\ 
\label{piw} 
\langle P\rangle_{\rm CL}&=& 0,\ \ \  \langle P^2\rangle_{\rm CL}=1. 
\end{eqnarray} 
So, we obtain the variances of $X$, $P$ as,  $(\Delta X)_{\rm CL}^2=\frac{1}{3}$ and 
 $(\Delta P)_{\rm CL}^2= 1$ for the classical ensemble of particles of fixed energy $E$, confined within the infinite well, 
which results in the following variance product: 
\begin{equation}
\label{uniw}
 (\Delta X)_{\rm CL}^2\, (\Delta P)_{\rm CL}^2\equiv\frac{1}{3}. 
\end{equation}

Quantum mechanical stationary state solutions (even and odd parity) for a particle confined in an one dimensional infinite potential well  (\ref{infpot}) are given by, 
\begin{eqnarray}
\label{iwq}
\psi^{(+)}_n(x)&=&\sqrt{\frac{2}{L}}\, \cos(n\,\pi\, x/L),\ \  n=1,3,5,\ldots  \nonumber \\ 
\psi^{(-)}_n(x)&=&\sqrt{\frac{2}{L}}\, \sin(n\,\pi\, x/L),\ \  n=2,4,6,\ldots \ 
\end{eqnarray}  
and the corresponding energy eigen values are, 
\begin{equation}
E_n=\frac{n^2\, \pi^2\, \hbar^2}{2\, m\, L^2}
\end{equation}
The scaled dimensionless position and momentum operators (analogous to the classical ones (\ref{iwXP})) are chosen as, 
\begin{eqnarray} 
\hat{X}=\frac{\hat{x}}{L/2},\ \ \hat{P}=\frac{\hat{p}}{\sqrt{2\,m\,E_n}}= \frac{ \hat{p}}{n\pi\hbar/L}.
\end{eqnarray} 
The expectation values of $\hat{X},\, \hat{X}^2, \hat{P}$, and $\hat{P}^2$ are evaluated in the stationary states (both even and odd) to obtain,
\begin{eqnarray}
\label{qiwx}
\langle \hat{X}\rangle_{QM}&=&\frac{1}{L/2}\, \int_{-L/2}^{L/2} \, dx\, \vert \psi^{\rm (+/-)}_n(x)\vert^2\, x =0 \\
\label{qiwxsq}
\langle \hat{X^2}\rangle_{QM}&=&\frac{1}{L^2/4}\,\, \int_{-L/2}^{L/2} \, dx\, \vert \psi^{\rm (+/-)}_n(x)\vert^2\, x^2  \nonumber \\
&=& \frac{1}{3}-\frac{2}{ n^2\,\pi^2}, \\
\label{qiwp}
\langle \hat{P}\rangle_{QM}&=&-i\frac{L}{n\pi}\, \int_{-L/2}^{L/2} \, dx\,  \psi^{(+/-)}_n(x)\frac{d\psi^{(+/-)}_n(x)}{dx} \nonumber \\ 
&=&0, \\
\langle \hat{P^2}\rangle_{QM}&=&-\frac{L^2}{n^2\pi^2}\, \int_{-L/2}^{L/2} \, dx\,  \psi^{(+/-)}_n(x)\frac{d^2\psi^{(+/-)}_n(x)}{dx^2}\nonumber \\ 
&=&1
\end{eqnarray}

It may be seen that $\langle \hat{X}^2\rangle_{\rm QM}\rightarrow \langle X^2\rangle_{\rm CL}=\frac{1}{3}$ in the large $n$ limit, in which case the  uncertainty product 
\begin{equation}
\label{qiwun} 
\lim_{n\rightarrow \infty}\, \left(\Delta\, \hat{X}\right)_{\rm QM}\,\left(\Delta\, \hat{P}\right)_{\rm QM}\approx \frac{1}{3}.  
\end{equation} 

From the commutator relation, 
\begin{eqnarray}
[\hat{X},\hat{P}]&=&\left[\frac{2}{L}\, \hat{x}, \frac{L}{n\pi\hbar}\, \hat{p}\right] \nonumber \\ 
&=&\frac{2\,i}{n\, \pi},
\end{eqnarray} 
it is clear that 
\begin{eqnarray}
 \left(\Delta\, \hat{X}\right)_{QM}^2\, \left(\Delta\, \hat{P}\right)_{QM}^2\geq \frac{1}{n^2\pi^2},
\end{eqnarray}
and in the large $n$ limit one recovers the expected result $\left(\Delta\, \hat{X}\right)_{QM}^2\, \left(\Delta\, \hat{P}\right)_{QM}^2\geq 0$. However, it may be noted that the stationary state uncertainty product (\ref{qiwun}) does not vanish in the limit $n\rightarrow \infty$ --    but it approaches the value $\frac{1}{3}$ -- which coincides {\em exactly} with that  associated with the classical ensemble (see (\ref{uniw})).     

\subsection{Bouncing ball} 

We now consider the example of a particle bouncing vertically  up and down in a uniform gravitational field, which is described by the confining potential, 
\begin{equation}
\label{bbpot}
V(z)=\left\{\begin{array}{ll} \infty & {\rm for}\  z<0 \\ 
mgz & {\rm for}\ \  z \geq 0.\end{array}\right. .
\end{equation} 
A classical particle of mass $m$, energy $E$ is under the influence of a constant force $F=-\frac{d V}{dz}=-mg,\ \ z\geq 0$ and it bounces back and forth  between $0\leq z\leq A$, where $A=E/mg$ denotes the maximum height reached.   

An ensemble of bouncing balls of energy $E$ is characterized by the classical position probability distribution~\cite{Rob}   
\begin{equation}
P_{\rm CL}(z)=\left\{\begin{array}{ll} \frac{1}{2\, A}\, \frac{1}{\sqrt{1-(z/A)}} & {\rm for} \ \ 0\leq z\leq A \\ 
0 & {\rm otherwise}. \end{array} \right. 
\end{equation}           
which is readily identified by substituting (\ref{bbpot}) in (\ref{posprob2})).

Employing dimensionless position and momentum variables 
\begin{equation}
\label{bbZP}
Z=\frac{z}{A}, \ \ P=\frac{p}{\sqrt{2mE}}=\frac{p}{\sqrt{2m^2\,g\, A}} 
\end{equation} 
(so that  $0\leq Z\leq 1$ and $-1\leq P\leq 1$ for the bouncing particles) we obtain the classical moments $\langle Z\rangle_{\rm CL},\ 
\langle Z^2\rangle_{\rm CL},\ \langle P\rangle_{\rm CL},\ \langle P^2\rangle_{\rm CL}$  as, 
\begin{eqnarray}
\label{zbb}
\langle Z\rangle_{\rm CL}&=& \frac{1}{A}\, \int\, dz\, P_{\rm CL}(z)\, z \nonumber \\ 
&=&\frac{1}{2\, A^2}\, \int_{0}^{A}\,  dz\, \frac{z}{\sqrt{1-(z/A)}} =\frac{2}{3}, \\ 
\label{zsqbb}
\langle Z^2\rangle_{\rm CL}&=&\frac{1}{A^2}\, \int\, dz\, P_{\rm CL}(z)\, z^2 \nonumber \\ 
&=&\frac{1}{2\, A^3 }\, \int_{0}^{A}\,  dz\, \frac{z^2}{\sqrt{1-(z/A)}} = \frac{8}{15},  \\ 
\label{pbb} 
\langle P\rangle_{\rm CL}&=& \frac{1}{2\sqrt{2m^2\, g\, A}}\, \int_{0}^{A}\, dz\,  P_{\rm CL}(z)\,
(-\sqrt{2m(E-mgz)}\nonumber \\ 
&& \ \ \ +\sqrt{2m(E-mgz)} = 0, \\
\label{psqbb} 
\langle P^2\rangle_{\rm CL}&=& \frac{1}{2m^2\, g\, A}\, \int_{0}^{A}\, dz\,  P_{\rm CL}(z)\,2m\, (E-mgz)\nonumber \\ 
& =&  \frac{1}{2\, A}\, \int_{0}^{A}\,  dz\, {\sqrt{1-(z/A)}} = \frac{1}{3}. 
\end{eqnarray} 
Thus, the variances of $Z$ and $P$ are given by     $(\Delta Z)_{\rm CL}^2=\frac{4}{45}$ and  
 $(\Delta P)_{\rm CL}^2=\frac{1}{3}$, 
leading to  
\begin{equation}
\label{unbb}
 (\Delta Z)_{\rm CL}^2\, (\Delta P)_{\rm CL}^2\equiv\frac{4}{135}. 
\end{equation}

Stationary state solutions of a quantum bouncer~\cite{Gea} (confining potential of which is given by (\ref{bbpot})) are obtained by solving the time-independent Schrodinger equation 
\begin{equation}
\label{schbb}
-\frac{\hbar^2}{2m}\frac{d^2\psi_n(z)}{dz^2}+mgz\, \psi_n(z)=E_n\, \psi_n(z), 
\end{equation}
with the boundary condition 
\begin{equation}
\label{bou}
\psi_n(0)=0.
\end{equation} 
 In terms of the characteristic {\em gravitational length}~\cite{Gea}
\begin{equation}
\label{lgbb}
l_g=\left(\frac{\hbar^2}{2\,m^2\, g}\right)^{1/3}, 
\end{equation} 
it is convenient to define dimensionless quantities 
\begin{eqnarray}
\label{enzprimebb}
E_n'=\frac{E_n}{mgl_g},\ \ \  z'=\frac{z}{l_g}-E_n' 
\end{eqnarray} 
so that the Schordinger equation (\ref{schbb}) takes the standard form    
\begin{equation}
\label{schbb2}
\frac{d^2\psi_n(z')}{dz'^2}=z'\, \psi_n(z') 
\end{equation}
which is the Airy differential equation. The solutions of (\ref{schbb2}) are two linearly independent sets of Airy functions $Ai(z'), Bi(z')$; however, the function $Bi$ diverges as its argument increases, and so it is not a physically admissible solution. The stationary state solutions of a quantum bouncer are thus given by,  
\begin{equation} 
\label{bbAi}
\psi_n(z')=N_n\, Ai(z'),\ \ z'\geq -E_n',
\end{equation} 
where $N_n$ denotes the normalization constant.  From the boundary condition (\ref{bou}), one obtains $Ai(-E_n')=0,\ \ n=1,2,\ldots$ leading to the identification that the (scaled) energy eigenvalues $E_n'$ are the $n$th zeros of the Airy function.  The first few energy eigenvalues $E_n'$ (first few zeros of  the Airy function) of the quantum bouncing ball are given in Table 1.

\begin{table}
\bigskip

\begin{tabular}{c| c}
\hline\hline 
 \hskip 0.5in $n\, \hskip 0.5in$  &  \hskip 0.5in $E_n'\, \hskip 0.5in$ \\ 
\hline\hline
1 & 2.3381 \\
\hline 
2 & 4.0879 \\
\hline 
3& 5.5205 \\
\hline  
4 & 6.7867 \\
\hline  
5 & 7.9441 \\
\hline\hline 
\end{tabular}
\caption{The first few scaled energy eigenvalues $E_n'$ of quantum bouncing ball.}
\end{table}    

\bigskip

Identifying the classical turning point $A_n$ associated with the energy eigenvalues $E_n$ of the quantum bouncer to be 
\begin{equation}
A_n=\frac{E_n}{mg}=l_g\, E'_n 
\end{equation}
we define appropriately scaled position and momentum operators (which are quantum counterparts of $Z,\ P$  defined in (\ref{bbZP})) as,    
\begin{eqnarray}
\label{qZPbb}
\hat{Z}&=&\frac{\hat{z}}{A_n}=\frac{\hat{z}}{l_g\, E'_n} \\ 
\hat{P}&=&\frac{\hat{p}}{\sqrt{2mE_n}}=\frac{l_g\, \hat{p}}{\hbar\, \sqrt{E'_n}}.\nonumber
\end{eqnarray}

Further, substituting (\ref{lgbb}), (\ref{enzprimebb}) in (\ref{qZPbb}), we may express the configuration representation of the operators  $\hat{Z}, \hat{P}$ in terms of $z', \ E_n'$ as follows:
\begin{eqnarray}
\hat{Z}&\rightarrow& \frac{1}{E_n'}\, (z'+E'_n) \\
\hat{P}&\rightarrow & \frac{-i}{\sqrt{E'_n}}\,\frac{\partial}{\partial\, z'}
\end{eqnarray}
The expectation values $\langle \hat{Z}\rangle_{QM}, \langle \hat{Z}^2\rangle_{QM}, \langle \hat{P}\rangle_{QM}, \langle \hat{P}^2\rangle_{QM}$ are then  evaluated (numerical computation using Mathematica package) in the eigen states (\ref{bbAi}) of the quantum bouncing ball:     
\begin{widetext}
\begin{eqnarray}
\label{qbbz}
\langle \hat{Z}\rangle_{QM}&=&\frac{1}{E'_n}\, \int_{-E'_n}^{\infty} \, dz'\, \vert \psi_n(z')\vert^2\, (z'+E'_n) \nonumber \\
&=& \frac{N^2_n}{E'_n}\, \int_{-E'_n}^{\infty} \, dz'\,  Ai^2(z')\, (z'+E'_n) =\frac{2}{3}  \\
\label{qbbzsq}
\langle \hat{Z}^2\rangle_{QM}&=&\frac{1}{E^{'2}_n}\, \int_{-E'_n}^{\infty} \, dz'\, \vert \psi_n(z')\vert^2\, (z'+E'_n)^2 \nonumber \\
&=& \frac{N^2_n}{E^{'2}_n}\, \int_{-E'_n}^{\infty} \, dz'\,  Ai^2(z')\, (z'+E'_n)^2 =\frac{8}{15}  
\end{eqnarray}

\begin{eqnarray}
\label{qbbp}
\langle \hat{P}\rangle_{QM}&=&\frac{-i}{\sqrt{E'_n}}\, \int_{-E'_n}^{\infty} \, dz'\, \psi^*_n(z')\, \frac{d\psi_n(z')}{dz'}  \nonumber \\
&=&\frac{-i\, N_n^2}{\sqrt{E'_n}}\, \int_{-E'_n}^{\infty} \, dz'\, Ai(z')\, \frac{d\, Ai(z')}{dz'} =0  \\
\label{qbbpsq}
\langle \hat{P}^2\rangle_{QM}&=&-\frac{1}{E'_n}\, \int_{-E'_n}^{\infty} \, dz'\, \psi^*_n(z')\, \frac{d^2\psi_n(z')}{dz^{'2}}  \nonumber \\
&=&-\frac{N_n^2}{E'_n}\, \int_{-E'_n}^{\infty} \, dz'\, Ai(z')\, \frac{d^2\, Ai(z')}{dz^{'2}} =\frac{1}{3}.  
\end{eqnarray}
\end{widetext}  
The quantum expectation values  (\ref{qbbz})-(\ref{qbbpsq}) in the eigen states  match identically with those (see (\ref{zbb})-(\ref{psqbb})) associated with a classical ensemble of bouncing balls. This is a novel identification, bringing forth the deep rooted unifying features in classical and quantum realms.   

From Eqs.~(\ref{qbbz})-(\ref{qbbpsq}), we obtain the variances of $\hat{Z}$ and $\hat{P}$ in the stationary states to be  $(\Delta \hat{Z})_{\rm QM}= 4/45$, $(\Delta \hat{P})_{\rm QM}= 1/3$ and hence, the uncertainty product  
\begin{equation}
\label{qunbb}
(\Delta \hat{Z})^2_{\rm QM}\, (\Delta \hat{P})^2_{\rm QM}\equiv \frac{4}{135},
\end{equation} 
which matches exactly with that of the classical ensemble of bouncing balls (see (\ref{unbb})).
It may be noted that, the commutation relation 
\begin{equation}
[\hat{Z},\hat{P}]=\left[\frac{\hat{z}}{l_g\, E'_n},\frac{l_g\, \hat{p}}{\hbar\, \sqrt{E'_n}}\right]=\frac{i}{(E'_n)^{3/2}}
\end{equation}
would lead to the uncertainty relation $(\Delta \hat{Z})^2_{\rm QM}\, (\Delta \hat{P})^2_{\rm QM}\geq \frac{1}{4\, (E'_n)^3}$. In the large $n$ limit $\frac{1}{E_n'}\rightarrow 0$  (as the energy eigenvalues obey the scaling relation~\cite{Rob}  $E'_n\propto n^{2/3}$ with $n$), thus resulting in     
the {\em classical limit} on the variance product  $(\Delta \hat{Z})^2_{\rm QM}\, (\Delta \hat{P})^2_{\rm QM}\geq 0$. The exact result on the uncertainties (see (\ref{qunbb})), which holds for all the eigenstates of the quantum bouncing ball, approaches the value $\frac{4}{135}$  for all values of $n$ and it agrees perfectly with that associated with the classical ensemble. 

\section{Summary}

Pedagogic discussions often attribute the emergence of classical mechanics from quantum theory to the limit  $\hbar\rightarrow 0$ and also, to that   based on Eherenfest's theorem. However, both these descriptions on classical regime are found to be unsatisfactory~\cite{Berry, Nemes, Ballentine1,Sen, Ballentine2, Angelo1, Angelo2}. Quantum predictions being purely statistical in nature, it is pertinent to compare the classical limit of  the quantum system with the corresponding classical ensemble -- not with a single particle~\cite{Ballentine1, Huang}. And a suitable {\em operational criterion} is therefore to compare the averages and probability distributions in both realms. In this connection, we have explored parallels between variances of dimensionless position and momentum variables -- and hence the resulting uncertainty relations --   in the stationary eigen states of one dimensional bound  quantum systems and their classical counterparts. In the specific examples of  harmonic oscillator, infinite well and  the bouncing ball, we have shown that the first and second moments of the scaled canonical observables {\em exactly} agree with each other -- resulting in identical  uncertainty relations in both quantum and classical regimes. This identification reflects a deep underlying connectivity between the two formalisms -- despite their differences in the mathematical and conceptual basis. 

\section*{Acknowledgement} 

We thank Professor A. K. Rajagopal and Dr. Sudha Shenoy for insightful comments.

\end{document}